# Understanding Opportunities and Risks of Synthetic Relationships: Leveraging the Power of Longitudinal Research with Customised AI Tools.

**Alfio Ventura** (corresponding author)[1][0000-0003-1639-8001] and Nils Köbis[1,2][0000-0002-0608-7333]

1 Research Center Trustworthy Data Science and Security, University of Duisburg-Essen, Bismarckstraße 120, 47057 Duisburg, Germany
2 Center for Humans and Machines, Max Planck Institute for Human Development, Lentzeallee 94, 14195 Berlin, Germany
[alfio.ventura;nils.koebis]@uni-due.de

**Abstract.** This position paper discusses the benefits of longitudinal behavioural research with customised AI tools for exploring the opportunities and risks of synthetic relationships. Synthetic relationships are defined as "continuing associations between humans and AI tools that interact with one another wherein the AI tool(s) influence(s) humans' thoughts, feelings, and/or actions." [1]. These relationships can potentially improve health, education, and the workplace, but they also bring the risk of subtle manipulation and privacy and autonomy concerns. To harness the opportunities of synthetic relationships and mitigate their risks, we outline a methodological approach that complements existing findings. We propose longitudinal research designs with self-assembled AI agents that enable the integration of detailed behavioural and self-reported data.

## 1 Introduction

Interactions with AI systems have become integral to many people's daily lives. Accruing users at a record-breaking pace, millions of people now interact with large language models (LLMs) like ChatGPT, Claude, or Gemini. Besides such general-purpose language models, there is a growing trend towards LLM-based AI tools aiming to develop sustainable relationships with their users [2]. Prominent examples are Replika and the soon-to-be-released "Friend AI" necklace that promises to be "always listening" [3]. Enhanced memory functions have enabled these tools to have more dynamic, personal, and persistent conversations with their users compared to previous technologies. They enable people to form synthetic relationships (SRs), defined by Starke and colleagues [1] as "continuing associations between humans and AI tools that interact with one another wherein the AI tool(s) influence(s) humans' thoughts, feelings, and/or actions."

## 2      The Opportunities of Synthetic Relationships

The growing sophistication of AI tools in universal agents opens up vast opportunities for SRs across various domains. These relationships are perceived as highly personalised, available 24/7, offering a safe space for self-disclosure without fear of judgment [4]. As companions, they might ease loneliness [5] and promote well-being [6]. In mental health, AI-based therapists have shown promise in reducing symptoms of depression and distress [7].

In education, AI tutors can tailor learning experiences to individual students' needs, fostering cognitive and socio-emotional development [8]. This personalised support can improve engagement and learning outcomes, particularly for students who struggle in traditional settings. Similarly, AI managers can enhance productivity in professional settings by streamlining decision-making and administrative tasks, freeing human managers to focus on more strategic initiatives [9].

## 3      The Risks of Synthetic Relationships

However, SRs also pose severe risks alongside these opportunities, stimulating first discussions [1, 10]. These risks are primarily ethical and affect the current state of normative behaviour. Privacy concerns loom large as AI agents gather and store private and intimate personal data, often without explicit user consent. In a world increasingly reliant on digital trust, this vulnerability could corroborate technology company's power and lead to new forms of exploitation. One of these risks is manipulation [1]. AI agents, by design, can influence user behaviour—ranging from "harmless" personalised product recommendations to dramatic steering of political opinions. SRs may amplify the effects of generative AI, such as emotional dependence and individualised responses, rendering users particularly susceptible to such influences.

There is also growing concern about the impact on human autonomy [10]. As users rely more on AI companions and therapists for emotional or social guidance, they risk ceding control over their personal and professional decisions. Over time, this could erode self-agency, leading to an over-reliance on synthetic relationships that might diminish critical thinking and individual autonomy. SRs may also disrupt social norms and reinforce maladaptive behaviours and cognitions by displaying sycophancy. Generative AI may create individualised echo chambers tailored towards the opinion of the interaction partner, reinforcing their beliefs and behavioural patterns.

SRs can also compromise our modern social networks. Their permanent availability, perceived safety and unjudgmental nature may make it difficult to argue in favour of human-human relationships over SRs in the future [10]. Ultimately, it is possible that "human relationships will just seem too hard" [11]. An interview study indicates that SRs use the same socialising resources as human-human relationships [4], challenging a future in which SRs are primarily supplementary. Replacing human-human relationships with SRs could be a natural consequence [12]. Recent news and discussions around a teen's suicide and the involvement of an SR illustrate how real the influence of SRs can already be today [13].

## 4 The Need for More Longitudinal Studies with Customized AI Agents

Despite the growing prevalence of SRs, the current body of research is limited in scope as most studies cannot detect the often subtle effects that unfold in synthetic relationships over time. Research has revealed the potential of AI tools to influence human feelings, attitudes, and behaviour. For instance, interview studies show that users often form attachment-like bonds with AI companions [14, 15]. This emotional bond can lead to adverse mental health effects [16]. Other studies found that interactions with opinionated AI assistants can sway participants' writing and opinions [17]. Moreover, AI-generated advice can affect people's ethical behaviour [18].

These studies provide important insights into immediate reactions to AI systems and can be regarded as snapshots of one-time interactions. This approach, however, has limitations. For one, such studies might suffer from the risks of experimenter demand effects as the interaction with the AI system is often (made) highly salient. Consequently, people might act in a way that they believe the experimenter wants them to. Also, short-term research designs are inadequate to assess many of the effects of sustained synthetic relationships. In fact, human-human relationship research suggests that understanding many important aspects of relationships requires a more extended scope of time as the effects are often more nuanced and take time to materialise [19, 20].

A few studies have already employed longitudinal research on ongoing user interactions with AI tools. One study examined how users' relationships with an AI companion evolve, showing that users' experiences and SR developments differ widely [21]. Another study found that AI companions significantly reduce loneliness on par with human interactions, especially when users feel "heard" by the chatbot [5].

These studies have started to uncover how attitudes towards and experiences with AI tools change over time. However, experimental research with proprietary software cannot provide transparent, reproducible and causal insights into how the design of AI systems in SRs affects people. Additionally, by relying on existing users of such tools, research may suffer from selection bias concerns limiting generalizability. Drawing on non-representative samples can significantly undermine quantitative behavioural insights, limiting the ability to systematically examine synthetic relationships' opportunities and risks. We, therefore, call for more research using a) customisable AI tools and b) implementing longitudinal designs, particularly with behavioural data.

### 4.1 Customizability for Causal Inferences

For independent, systematic research, control over the AI tool people interact with is paramount. Comprehensively prompted or fine-tuned AI tools offer several key advantages over relying on proprietary software. First, it provides high control over the backend, enabling researchers to tailor AI systems specifically for their studies. This control allows for precise adjustments in how the AI interacts with users, ensuring that variables of interest can be closely monitored and manipulated. Second, custom AI tools allow for the implementation of experimental treatments in ways that proprietary platforms often cannot support. Researchers can design multiple versions of the AI to test different features or interaction styles, thus providing causal insights into how specific design choices affect user behaviour and outcomes.

Furthermore, by developing AI tools internally, researchers eliminate potential selection biases inherent in existing platforms' user bases and ensure that study participants interact with tools designed precisely for the experimental conditions, thereby enhancing the validity and robustness of the findings. Finally, the researchers can transparently handle the user data with explicit participant consent and according to GDPR. Such tools and data should then be openly shared with the research community, ensuring replicability and reusability for everyone with adequate hardware.

Internally developed AI tools also allow the control of the adaptation process of the AI and thus address the reflection problem [22] for causal inferences. In human-AI interaction, the reflection problem emerges as both parties adapt to each other over time. Continuous adjustments by the AI make it challenging to discern whether the interactions with the AI cause observed changes in human behaviour or if both parties are merely adapting to a shared context. This mutual adaptation obscures whether AI's influence on the user is distinct from the user's impact on the AI's adjustments. The challenge is, therefore, to disentangle reciprocal effects from changes driven by external or contextual factors influencing both parties. Control over the AI tools allows researchers to control AI adaption processes, for example, in a staggered adjustment design, and infer causality. Instead of allowing the AI to adapt continuously, researchers may periodically adjust the AI's behaviour according to a randomised schedule, assigning groups of participants to experience delayed or limited AI adjustments at various intervals. By controlling the timing and frequency of these adjustments, researchers can isolate the influence of AI adjustments by comparing groups exposed to different adaptation schedules. Such an experimental design allows for comparing user behaviour between groups, helping to distinguish the effects of the AI's adaptive influence from changes stemming from the human side or shared contextual factors.

### 4.2     Longitudinal Quantitative (Behavioural) Data

Longitudinal quantitative studies of SRs offer a comprehensive understanding of how users' behaviours evolve over time. Behavioural measures, such as tracking individual interaction frequency or the duration and specific types of engagement (e.g., emotional disclosure and task completion), provide objective data beyond self-reported attitudes or perceptions. These measures can reveal longitudinal patterns in AI reliance, emotional dependency, and habit formation—offering insights that users may not consciously report or even be aware of. For example, users might not recognise an increasing reliance on AI for decision-making, but behavioural data could display this trend more objectively.

Behavioural data can also enrich qualitative insights by adding measurable depth to users' stated experiences. While qualitative studies can capture the emotional and cognitive nuances of SRs and inspire research questions, quantitative data can validate such findings. By integrating both approaches, researchers can correlate shifts in stated attitudes with actual behaviours, identifying discrepancies and refining our understanding of SR dynamics over time. Quantitative data also facilitates replication and generalisation, which are critical for building robust models of how SRs influence long-term human behaviour across diverse populations.

### 4.3 Experience Sampling

Experience sampling is one method that holds particular promise for longitudinal data collection. For a thorough explanation, the work of Csikszentmihalyi and Larson [23] highlights Experience Sampling (ESM)'s ability to track real-time experiences. Such momentary assessment technology captures developing SRs by tracking (self-reported) emotional responses, thoughts and behaviours during user interactions throughout the day. This rich data offers a nuanced qualitative and quantitative understanding of how SRs evolve. Illustrating the power of ESM, a systematic literature review shows that smartphone-based experience sampling is essential for understanding well-being dynamics [24]. Applied to SRs, ESM enables researchers to observe subtle shifts in trust, engagement, and dependency over extended periods, offering crucial insights into how these relationships shape long-term behaviours and attitudes. This strength holds true, even if such subtle changes may not be consciously acknowledged but emerge from ongoing interactions and relationship building.

## 5 Exemplary Research Agenda: Experience Sampling in Synthetic Relationships

A longitudinal study using behavioural methods combined with experience sampling could provide comprehensive insights into both the opportunities and risks of SRs. For example, research questions about patterns of emotional disclosure, frequency and motivation of companionship seeking, or calibration of emotional dependence and trust within SRs can be explored, possibly even with the same data set. By utilising experience sampling's comprehensive usage data, researchers can explore complex dyadic relationship patterns evolving over time. From there, they can draw educated recommendations for the socially beneficial application, design and boundaries of SRs. In the future, this approach could be extended to investigate SRs in hybrid human-AI social networks.

Two areas where this methodology could be particularly insightful are (1) assessing the risk of autonomy erosion in decision-making and (2) changes in users' social behaviours with and through AI. In terms of decision making, a combination of behavioural tracking and experience sampling could reveal how AI-driven advice influences users' actions, whether in making purchases, forming opinions, or navigating ethical dilemmas. Over time, such studies could show how AI agents subtly steer user behaviour, shifting the balance between human autonomy and AI-influenced behaviour.

Regarding social behaviour, monitoring interactions with both AI-human and human-human dyads could suggest whether long-term engagement with AI agents leads to a preference for synthetic interactions over human ones. For example, comprehensive long-term behavioural and emotional data may inform about changes in reliance on AI vs human partners for emotional or social support. Such changes in reliance may indicate a shift from human-human relationships to SRs, possibly accompanied by deteriorating social skills and social networks–a long-term shift in behaviour resulting from sustained AI interaction and SRs that may not be captured with other methods.

## 6      Conclusion

Longitudinal, quantitative behavioural research is indispensable to fully grasp the opportunities and risks of synthetic relationships. By leveraging customised AI tools and methods such as experience sampling, researchers can capture the complex, evolving dynamics of human-AI interactions over time. Behavioural data, which provides objective insights into user engagement, can enrich our understanding beyond self-reported attitudes, helping to uncover subtle shifts in emotional dependency, decision-making autonomy, and social behaviour. This approach will pave the way for more informed, evidence-based development of AI systems that are both innovative and ethically sound.

**Disclosure of Interests.** The authors have no competing interests to declare relevant to this article's content.